\documentclass[aps,prl,twocolumn,showpacs,preprintnumbers,amsmath,amssymb,floatfix,nofootinbib]{revtex4-1}
\usepackage{graphicx}

\usepackage{color}
\usepackage{colordvi}

\usepackage[normalem]{ulem}

\usepackage{color}

\newcommand{\beqn}{\begin{eqnarray}}
\newcommand{\eeqn}{\end{eqnarray}}
\newcommand{\be}{\begin{equation}}
\newcommand{\ee}{\end{equation}}

\newcommand{\ba}{\begin{array}{c}}
\newcommand{\bat}{\begin{array}{cc}}
\newcommand{\ea}{\end{array}}

\newcommand{\bi}{\begin{itemize}}
\newcommand{\ei}{\end{itemize}}

\newcommand{\ket}{\,\rangle}
\newcommand{\bra}{\langle \,}

\newcommand{\Frac}[2]{\frac{\displaystyle #1}{\displaystyle #2}}

\newcommand{\Int}{\displaystyle{\int}}

\begin{document}

\preprint{IFIC/12-87}
\preprint{FTUV/12-1228}


\title{Viability of strongly-coupled scenarios with a light Higgs-like boson}

\author{Antonio Pich${}^{1}$}
\author{Ignasi Rosell${}^{1,2}$}
\author{Juan Jos\'e Sanz-Cillero${}^{3,4}$}

\affiliation{${}^1$ Departament de F\'\i sica Te\`orica, IFIC, Universitat de Val\`encia -- CSIC, Apt. Correus 22085, 46071 Val\`encia, Spain }

\affiliation{${}^2$  Departamento de Ciencias F\'\i sicas, Matem\'aticas y de la Computaci\'on,
Universidad CEU Cardenal Herrera, 46115 Alfara del Patriarca, Val\`encia, Spain}

\affiliation{${}^3$  INFN -- Sezione di Bari, Via Orabona 4, 70126 Bari, Italy }

\affiliation{${}^4$  Department of Physics, Peking University, Beijing 100871, P.R. China }

\begin{abstract}

We present a one-loop calculation of the oblique $S$ and $T$ parameters within
strongly-coupled models of electroweak symmetry breaking with a light Higgs-like boson.
We use a general  effective Lagrangian,
implementing the chiral symmetry breaking $SU(2)_L\otimes SU(2)_R\to SU(2)_{L+R}$ with Goldstones, gauge bosons,
the Higgs-like scalar and one multiplet of vector and axial-vector massive resonance states.
Using a dispersive representation and imposing a proper ultraviolet behaviour,
we obtain $S$ and $T$ at the next-to-leading order in terms of a few resonance parameters.
The experimentally allowed range forces the vector and axial-vector states to be heavy, with masses
above the TeV scale, and suggests that the Higgs-like scalar should have a $WW$ coupling close to the Standard Model one.
Our conclusions are generic and apply to more specific scenarios such as
the minimal $SO(5)/SO(4)$ composite Higgs model.

\end{abstract}



\maketitle

\vspace{-5cm}

\section{Introduction}

A new Higgs-like boson around $126\,$GeV has just been discovered at the LHC~\cite{LHC}.
Although its properties are not well measured yet, it complies with the expected behaviour
and therefore it is a very compelling candidate to be the Standard Model (SM) Higgs.
An obvious question to address is to which extent alternative scenarios of
Electroweak Symmetry Breaking (EWSB) can be already discarded or strongly constrained.
In particular, what are the implications for strongly-coupled models where the
electroweak symmetry is broken dynamically?

The existing phenomenological tests have confirmed the
$SU(2)_L\otimes SU(2)_R\rightarrow SU(2)_{L+R}$
pattern of symmetry breaking, giving rise to three Goldstone bosons which, in the
unitary gauge, become the longitudinal polarizations of the gauge bosons.
When the $U(1)_Y$ coupling $g'$ is neglected, the electroweak Goldstone dynamics
is described at low energies by the same Lagrangian as
the QCD pions, replacing the pion decay constant by the
EWSB scale $v=(\sqrt{2}G_F)^{-1/2} = 246\,$GeV~\cite{AB:80}.
Contrary to the SM, in strongly-coupled scenarios the symmetry is nonlinearly realized and one expects the appearance of massive resonances generated by the non-perturbative interaction.

The dynamics of Goldstones and massive resonance states can be analyzed in a generic way by using
an effective Lagrangian, based on symmetry considerations.
The theoretical framework is completely analogous to the Resonance Chiral Theory description
of QCD at GeV energies~\cite{RChT}.
Using these techniques, we investigated in Ref.~\cite{paper}  the oblique $S$ parameter~\cite{Peskin:92}, characterizing the
electroweak boson self-energies, within Higgsless strongly-coupled models.
Adopting a dispersive approach and imposing a proper UV behaviour, it was shown there that it is possible
to calculate $S$ at the next-to-leading order, {\it i.e.}, at one-loop.
We found that in most strongly-coupled scenarios of EWSB a high resonance mass scale is required, $M_V > 1.8\,$TeV,
to satisfy the stringent experimental limits.

The recent discovery of a Higgs-like boson makes mandatory to update the analysis, including the light-scalar contributions.
In addition, we will also present a corresponding one-loop calculation of the oblique $T$ parameter, which allows us to perform
a correlated analysis of both quantities. $S$ measures the difference
between the off-diagonal $W^3B$ correlator and its SM value, while $T$ parametrizes
the breaking of custodial symmetry~\cite{Peskin:92}.
More precisely, $T$ measures the difference between the $W^3W^3$ and $W^+W^-$ correlators,
subtracting the SM contribution.
The explicit definitions of $S$ and $T$ are given in Refs.~\cite{Peskin:92,paper}.
Previous one-loop analyses
can be found in Refs.~\cite{other,S-Orgogozo:11,Orgogozo:2012}.

\section{Theoretical Framework}

We have considered a low-energy effective theory containing the SM gauge bosons coupled
to the electroweak Goldstones, one light scalar state $S_1$ with mass $m_{S_1} = 126$~GeV
and the lightest vector and axial-vector resonance multiplets $V_{\mu\nu}$ and $A_{\mu\nu}$.
We only assume the SM pattern of EWSB, {\it i.e.} the theory
is symmetric under $SU(2)_L\otimes SU(2)_R$ and becomes spontaneously broken to the diagonal
subgroup $SU(2)_{L+R}$.
$S_1$ is taken to be singlet under $SU(2)_{L+R}$, while $V_{\mu\nu}$ and $A_{\mu\nu}$ are triplets (singlet vector and axial-vector contributions are absent at the order we are working).
To build the Lagrangian we only consider operators with the lowest number of derivatives,
as higher-derivative terms are either proportional to the equations of motion or tend to violate the expected short-distance behaviour of the Green's functions~\cite{RChT}.
We will need the interactions
\begin{eqnarray}\label{eq:Lagrangian}
\mathcal{L}\; &=\; &
\frac{v^2}{4}\,\bra \! u_\mu u^\mu \!\ket\,\left( 1 + \frac{2\,\omega}{v}\, S_1\right)
 + \frac{F_A}{2\sqrt{2}}\, \bra \!A_{\mu\nu} f^{\mu\nu}_- \!\ket
\nonumber \\ &&\mbox{}
+ \frac{F_V}{2\sqrt{2}}\, \bra \!V_{\mu\nu} f^{\mu\nu}_+ \!\ket
+ \frac{i\, G_V}{2\sqrt{2}}\, \bra \! V_{\mu\nu} [u^\mu, u^\nu] \!\ket
\nonumber \\ &&\mbox{}
+ \sqrt{2}\, \lambda_1^{SA}\,  \partial_\mu S_1 \, \bra \! A^{\mu \nu} u_\nu \!\ket\, ,
\end{eqnarray}
plus the standard gauge boson and resonance kinetic terms.
The three Goldstone fields $\vec\pi(x)$ are parametrized through the matrix
$U=u^2= \exp{\left\{ i \vec{\sigma} \vec{\pi} / v \right\} }$,
$u^\mu = -i\, u^\dagger  D^\mu U\, u^\dagger$ with $D^\mu$
the appropriate gauge-covariant derivative, and
$\langle A\rangle$ stands for the trace of the $2\times 2$ matrix $A$. We follow the notation from Ref.~\cite{paper}.
The first term in (\ref{eq:Lagrangian}) gives the Goldstone Lagrangian, present in the SM, plus the
scalar-Goldstone interactions.
For $\omega=1$ one recovers the $S_1\to\pi\pi$ vertex of the SM. The calculation
will be performed in the Landau gauge, which eliminates the mixing between Goldstones and  gauge bosons.

The oblique parameter $S$ receives tree-level contributions from vector and axial-vector exchanges \cite{Peskin:92}, while
$T$ is identically zero at lowest-order (LO):
\begin{equation}
S_{\mathrm{LO}} = 4\pi \left( \frac{F_V^2}{M_V^2}\! -\! \frac{F_A^2}{M_A^2} \right)  \,,
\qquad\quad
T_{\mathrm{LO}}=0 \,.
\label{eq:LO}
\end{equation}
To compute the one-loop contributions we use the dispersive representation of $S$ introduced
by Peskin and Takeuchi~\cite{Peskin:92},
whose convergence requires a vanishing spectral function at short distances:
\begin{equation}
S\, =\, \Frac{16 \pi}{g^2\tan\theta_W}\,
\Int_0^\infty \, \Frac{{\rm dt}}{t} \, [\, \rho_S(t)\, - \, \rho_S(t)^{\rm SM} \, ]\, ,
\end{equation}
with $\rho_S(t)\,\,$ the spectral function of the $W^3B$ correlator~\cite{paper,Peskin:92}.
We work at lowest order in $g$ and $g'$ and only the lightest
two-particle cuts have been considered, {\it i.e.} two Goldstones or one Goldstone plus one scalar
resonance. $V\pi$ and $A\pi$ contributions were shown to be very suppressed in Ref.~\cite{paper}.

The calculation of $T$  is simplified by noticing that, up to corrections of $\mathcal{O}(m_W^2/M_R^2)$,
\begin{equation}
\alpha\,
T\,=\,\frac{Z^{(+)}}{Z^{(0)}}-1 \,,
\end{equation}
where $Z^{(+)}$ and $Z^{(0)}$ are the wave-function renormalization constants of the charged
and neutral Goldstone bosons computed in the Landau gauge~\cite{Barbieri:1992dq}.
A further simplification occurs by setting to zero $g$, which does not break the custodial symmetry, so
only the $B$-boson exchange produces an effect in $T$.
This approximation captures the lowest order contribution to $T$   in its expansion
in powers of   $g$ and $g'$.
Again only the lowest two-particle
cuts have been considered, {\it i.e.} the $B$ boson plus one Goldstone or one scalar resonance.

\begin{figure}
\begin{center}
\includegraphics[scale=0.27]{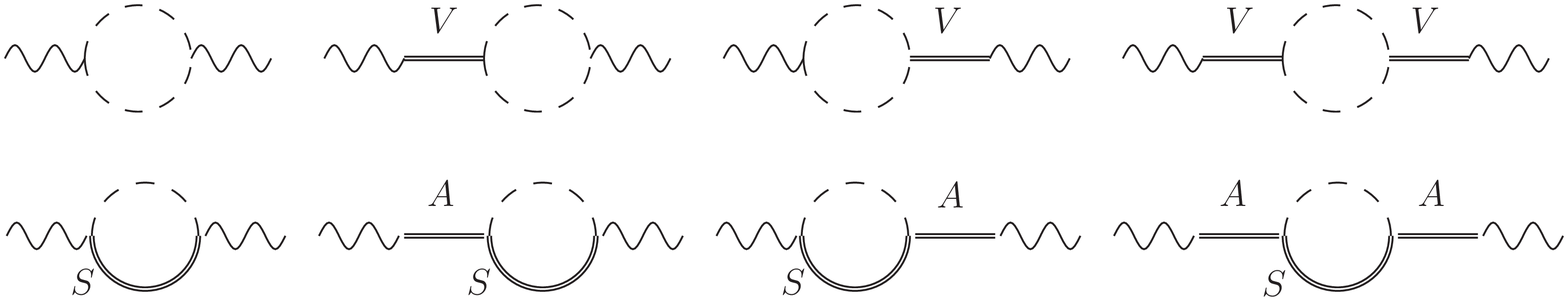}
\\[8pt]
\includegraphics[scale=0.27]{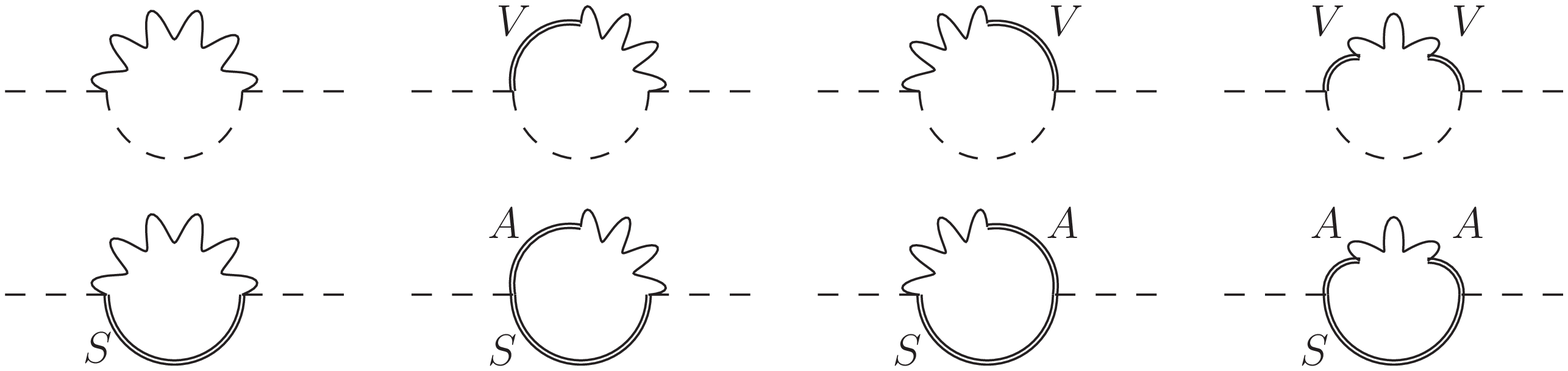}
\caption{\small{NLO contributions to $S$ (two first lines) and $T$ (two last lines).
A dashed (double) line stands for a Goldstone (resonance) boson and a curved line represents a gauge boson.}}
\label{NLO_graphs}
\end{center}
\end{figure}

Fig.~\ref{NLO_graphs} shows the computed one-loop contributions to $S$ and $T$.
Requiring the
$W^3 B$ correlator to vanish at high energies implies also a good convergence of the Goldstone
self-energies, at least for the two-particle cuts we have considered.
Therefore their difference obeys an unsubtracted
dispersion relation, which enables us to also compute $T$ through the dispersive integral,
\begin{eqnarray}
T &=& \Frac{4 \pi}{g'^2 \cos^2\theta_W}\, \Int_0^\infty \,\Frac{{\rm dt}}{t^2} \,
[\, \rho_T(t) \, -\, \rho_T(t)^{\rm SM} \,] \, ,
\end{eqnarray}
with $\rho_T(t)\,\,$
 the spectral function of the difference of the neutral and charged Goldstone self-energies.

\section{Short-distance constraints}

Fixing the scalar mass to $m_{S_1}=126$~GeV, we have 7 undetermined parameters: $M_V$, $M_A$, $F_V$, $G_V$, $F_A$, $\omega$ and $\lambda_1^{SA}$.
The number of unknown couplings can be reduced using short-distance information.

Assuming that weak isospin and parity are good symmetries of the strong dynamics, the $W^3 B$ correlator
is proportional to the difference of the vector and axial-vector two-point Green's functions~\cite{Peskin:92}.
In asymptotically-free gauge theories this difference vanishes at $s\to\infty$ as $1/s^3$ \cite{Bernard:1975cd}, implying two super-convergent sum rules,
known as the first and second Weinberg sum rules (WSRs)~\cite{WSR}, which at LO give the relations
\begin{equation}
F_{V}^2 - F_{A}^2  = v^2\, ,\qquad F_{V}^2  \,M_{V}^2 - F_{A}^2 \, M_{A}^2  = 0\, .
\end{equation}
This determines $F_V$ and $F_A$ in terms of the resonance masses, leading to
\begin{equation}
S_{\mathrm{LO}}\; =\; \frac{4\pi v^2}{M_V^2}\,  \left( 1 + \frac{M_V^2}{M_A^2} \right) \, .
\end{equation}
Since the WSRs also imply $M_A>M_V$, this prediction turns out to be bounded by~\cite{paper}
\begin{equation}
\frac{4\pi v^2}{M_V^2} \; < \; S_{\rm   LO} \; < \;   \frac{8 \pi v^2}{M_V^2} \, . \label{SLOtwoWSR}
\end{equation}

It is likely that the first WSR is also true in gauge theories
with non-trivial ultraviolet fixed points~\cite{S-Orgogozo:11,WalkingTC}  whilst
the second WSR is questionable in some scenarios.
If only the first WSR is considered,
but still assuming the hierarchy  $M_A>M_V$,  one obtains the lower bound~\cite{paper}
\begin{equation}
S_{\mathrm{LO}}
= 4\pi \left\{ \frac{v^2}{M_V^2} + F_A^2 \left( \frac{1}{M_V^2} - \frac{1}{M_A^2} \right)
\right\}> \frac{4\pi v^2}{M_V^2}     .
\label{eq.LO-S+1WSR}
\end{equation}
The possibility of an inverted mass ordering of the vector and axial-vector resonances~\cite{WalkingTC}
would turn this lower bound into the upper bound
$S_{\mathrm{LO}} < 4\pi v^2/M_V^2 $. Note that if the splitting of the vector and axial-vector
resonances was small, the prediction of $S_{{\rm LO}}$ would be close to the bound.

At the next-to-leading order (NLO)
 the computed $W^3 B$ correlator should also satisfy the proper short-distance behaviour.
The $\pi\pi$ and $S\pi$ spectral functions would have an unphysical grow at large momentum transfer
unless
$F_V G_V = v^2$ and $F_A \lambda^{SA}_1 = \omega v$.
The first constraint guarantees a well-behaved vector form factor \cite{RChT}, while the second
relates the axial and scalar couplings. Once these relations are enforced, the Goldstone self-energies
are convergent enough to allow for an unambiguous determination of $T$ in terms of masses and
$\omega$.
Neglecting terms of $\mathcal{O}(m_{S_1}^2/M_{V,A}^2)$,
\begin{equation}
 T\; =\;  \frac{3}{16\pi \cos^2 \theta_W} \bigg[ 1 + \log \frac{m_{H}^2}{M_V^2}
 - \omega^2 \left( 1 + \log \frac{m_{S_1}^2}{M_A^2} \right)  \bigg]  \, ,
\label{eq:T}
\end{equation}
where $m_H$ is the SM reference Higgs mass adopted to define $S$ and $T$.
Notice that taking $m_H=m_{S_1}$ and $\omega = 1$ (the SM value), $T$
vanishes when $M_V=M_A$ as it should.

To enforce the second WSR at NLO one needs the additional constraint
$\omega = M_V^2/M_A^2$ (constrained to the range $0\leq \omega \leq 1$).
One can then obtain a NLO determination of $S$ in terms of $M_V$ and $M_A$:
\begin{eqnarray}
S \; &=\; &   4 \pi v^2 \left(\frac{1}{M_{V}^2}+\frac{1}{M_{A}^2}\right) + \frac{1}{12\pi}
\bigg[ \log\frac{M_V^2}{m_{H}^2}  -\frac{11}{6}
\nonumber  \\ &&
+\;\frac{M_V^2}{M_A^2}\log\frac{M_A^2}{M_V^2}
 - \frac{M_V^4}{M_A^4}\, \bigg(\log\frac{M_A^2}{m_{S_1}^2}-\frac{11}{6}\bigg) \bigg] \,,\quad
\label{eq.1+2WSR}
\end{eqnarray}
where terms of $\mathcal{O}(m_{S_1}^2/M_{V,A}^2)$ have been neglected.
Taking $m_H=m_{S_1}$, the correction to the LO result vanishes when $M_V=M_A$ ($\omega=1$); in this limit, the NLO prediction reaches the LO upper bound in Eq.~(\ref{SLOtwoWSR}).

If only the first WSR is considered, one can still obtain a lower bound at NLO in terms of
$M_V$, $M_A$ and $\omega$:
\begin{equation}
S \geq   \frac{4 \pi v^2}{M_{V}^2} + \frac{1}{12\pi}  \bigg[ \log\frac{M_V^2}{m_{H}^2} -\frac{11}{6}
- \omega^2 \bigg(\!\log\frac{M_A^2}{m_{S_1}^2}-\frac{17}{6}
 + \frac{M_A^2}{M_V^2}\!\bigg) \bigg]  ,
\label{eq.lower-bound-1WSR}
\end{equation}
where $M_V<M_A$  has been assumed and we have neglected again terms
of $\mathcal{O}(m_{S_1}^2/M_{V,A}^2)$.
With $m_H=m_{S_1}$, the NLO correction vanishes in the combined limit $\omega=1$ and $M_V=M_A$, where the LO lower bound (\ref{eq.LO-S+1WSR}) is recovered.

\section{Phenomenology}

\begin{figure}
\begin{center}
\includegraphics[scale=0.6]{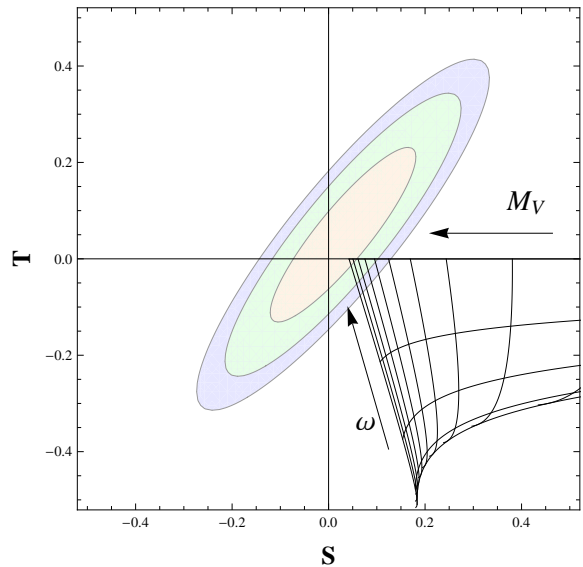}
\caption{\small{NLO determinations of $S$ and $T$, imposing the two WSRs.
The approximately vertical curves correspond to constant values
of $M_V$, from $1.5$ to $6.0$~TeV at intervals of $0.5$~TeV.
The approximately horizontal curves have constant values
of $\omega$:
$0.00, \, 0.25, 0.50, 0.75, 1.00$.
The arrows indicate
the directions of growing  $M_V$ and $\omega$.
The ellipses give the experimentally allowed regions at 68\% (orange), 95\% (green) and 99\% (blue) CL.}}
\label{fig.2WSR}
\end{center}
\end{figure}

Taking the SM reference point at $m_H = m_{S_1}= 126$ GeV,
the global fit to precision electroweak data gives the results
$S = 0.03\pm 0.10$ and $T=0.05\pm0.12$, with a correlation coefficient of $0.891$~\cite{phenomenology}.
In Fig.~\ref{fig.2WSR}
we show the compatibility between these ``experimental'' values and our NLO determinations
imposing the two WSRs: Eq.~(\ref{eq:T}) with
$\omega=M_V^2/M_A^2$ and Eq.~(\ref{eq.1+2WSR}).
Notice that the line with $\omega= M_V^2/M_A^2=1$ ($T=0$)
coincides with the LO upper bound in~(\ref{SLOtwoWSR}), while the
$\omega \,= M_V^2/M_A^2\to 0$
curve reproduces the lower bound in Eq.~(\ref{eq.lower-bound-1WSR})
in the same limit.
Thus, a vanishing scalar-Goldstone coupling ($\omega=0$) would be incompatible with the data,
independently of whether the second WSR has been assumed.

Fig.~\ref{fig.2WSR} shows a very important result in the two-WSR scenario: with $m_{S_1} = 126$~GeV, the precision electroweak data
requires that the Higgs-like scalar should have a $WW$ coupling very close to the
SM one. At 68\% (95\%) CL, one gets
$\omega\in [0.97,1]$  ($[0.94,1]$),
in nice agreement with the present LHC evidence \cite{LHC}, but much more restrictive.
Moreover, the vector and axial-vector states should be very heavy (and quite degenerate);
one finds $M_V> 5$~TeV ($4$~TeV) at 68\% (95\%) CL.

\begin{figure}
\begin{center}
\includegraphics[scale=0.65]{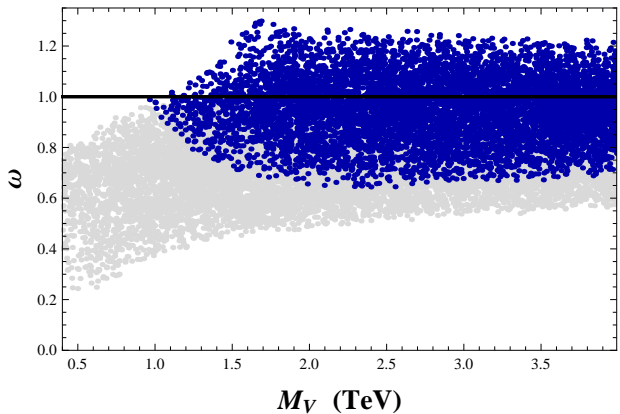}
\caption{\small Scatter plot for the 68\% CL region, in the case
when only the first WSR is assumed.
The dark blue and light gray regions
correspond, respectively,  to
$0.2<M_V/M_A<1$ and $0.02<M_V/M_A<0.2$.
}
\label{fig.1WSR}
\end{center}
\end{figure}

This conclusion is softened when the second WSR is dropped
and the lower bound in Eq.~(\ref{eq.lower-bound-1WSR}) is used instead.
This is shown in Fig.~\ref{fig.1WSR}, which gives
the allowed 68\% CL region in the space of parameters $M_V$ and $\omega$,
varying $M_V/M_A$ between 0 and 1. Note, however, that values of $\omega$
very different from the SM can only be obtained with a large
splitting of the vector and axial-vector masses. In general there is no solution for $\omega >1.3$.
Requiring $0.2<M_V/M_A<1$, leads to $1-\omega<0.4$
at 68\% CL,  while the allowed vector mass stays above 1~TeV~\cite{Filipuzzi:2012bv}.
Taking instead  $0.5<M_V/M_A<1$, one gets the stronger constraints
$1-\omega <0.16$ and $M_V>1.5$~TeV.
In order to allow vector masses below the TeV scale, one needs
a much larger resonance-mass splitting,
so that the NLO term in (\ref{eq.lower-bound-1WSR}) proportional to $\omega^2$ 
compensates the growing of the LO vector contribution.
The mass splitting gives also an additive contribution to $T$ of the form
$\delta T\sim \omega^2 \log{(M_A^2/M_V^2)}$,
making lower values of $\omega$ possible for smaller $M_V$.
However, the limit $\omega\to 0$ can only be approached
when $M_A/M_V\to \infty$.

In summary, strongly-coupled electroweak models with massive resonance states
are still allowed by the current experimental data.
Nonetheless, 
the recently discovered Higgs-like boson
with mass $m_{S_1}=126$~GeV must have a $WW$ coupling close to the SM one ($\omega=1$).
In those scenarios, such as asymptotically-free theories,
where the second WSR is satisfied, the $S$ and $T$ constraints force $\omega$ to be in the range
$\left [ 0.94, 1\right]$ at 95\% CL. Larger departures of the SM value can be accommodated when
the second WSR does not apply, but one needs to introduce a correspondingly large mass splitting between the vector and axial-vector states.

Similar conclusions can be obtained within more specific models, particularizing our general framework. For instance, let us mention the recent phenomenological analyses of vector and axial-vector states within the $SO(5)/SO(4)$ minimal composite Higgs model~\cite{Orgogozo:2012,composite}.
In this context, our scalar coupling would be
related to  the $SO(4)$ vacuum angle $\theta$ and upper bounded in the form
$\omega=\cos\theta\leq 1$~\cite{composite}. With this identification, 
the $S$ and $T$ constraints in Fig.~\ref{fig.2WSR} remain valid in this composite scenario.

\vspace{0.3cm}

\acknowledgments

{\bf Acknowledgments}: This work has been supported in part
by the Spanish Government [grants FPA2007-60323, FPA2011-23778,
CSD2007-00042 (Consolider Project CPAN)
and MICINN-INFN AIC-D-2011-0818], the Italian Government [grant MIUR-PRIN-2009], the National Nature Science Foundation of China [grant 0925522] and
the Universidad CEU Cardenal Herrera [grant PRCEU-UCH35/11].
J.J.S.C. thanks G. Cacciapaglia, H.Y. Cai and H.Q. Zheng for useful discussions.


\end{document}